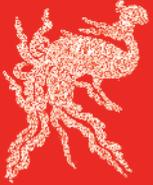

# Azerbaijan Journal of Physics

# Fizika

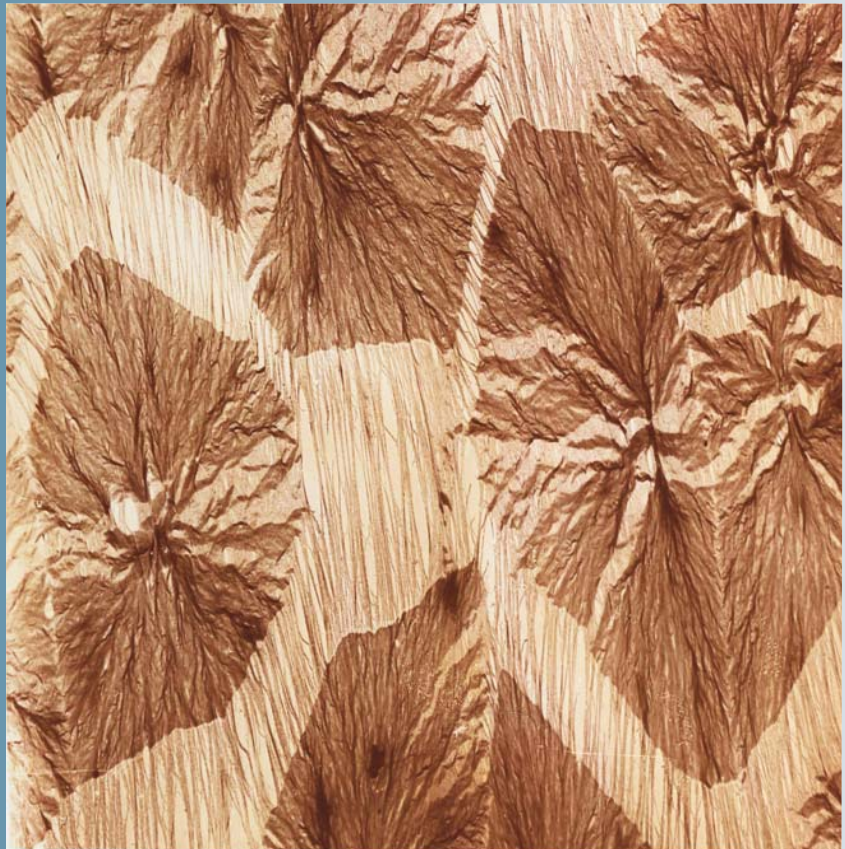

# NUCLEAR TRANSPARENCY EFFECT IN THE STRONGLY INTERACTING MATTER

M. AJAZ[1], M.K. SULEYMANOV[1,2], O.B. ABDINOV[3], ALI ZAMAN[1], K.H. KHAN[1], Z. WAZIR[1], Sh. KHALILOVA[3]

[1]COMSATS Institute of Information Technology, Department of Physics, Islamabad 44000, Pakistan
[2] Veksler and Baldin Laboratory of High Energy Physics, JINR, Dubna 141980, Russia
[3]H.M. Abdullayev Institute of Physics NAS Azerbaijan Republic, Baku

The work supported by the Science Development Foundation under the President of the Azerbaijan Republic.

We discuss that the results of study of the nuclear transparency effect in nuclear-nuclear collisions at relativistic and ultrarelativistic energies could help to extract the information on new phases of the strongly interacting matter as well as the *QCD* critical point. The results could provide further confirmation of the existence of the "horn" effect which had initially been obtained for the ratio of average values of $K^+$- to $\pi^+$-mesons' multiplicity as a function of the initial energies in the *NA49 SPS CERN* experiment. To observe the "horn" as a function of centrality, the new more enriched experimental data are required. The data which are expected from *NICA/MPD JINR* and *CBM GSI* setups could fulfill the requirement.

**PACS:** 539.12/.17

**INTRODUCTION**

Search for the new phases of strongly interacting matter as well as the Quantum Chromodynamics (*QCD*) critical point is one of the main objectives of the modern nuclear physics at high energies. Ultrarelativistic heavy ion collisions provide a unique opportunity to create and study the nuclear matter at high densities and temperatures. The produced state will pass different phases of the strongly interacting matter. The effects of: dilepton production; thermal radiation; strangeness enhancement; $J/\Psi$ suppression; jet suppression; flow are considered as possible signatures on the phases of the strongly interacting matter as well as the *QCD* critical point and Quark Gluon Plasma (*QGP*).

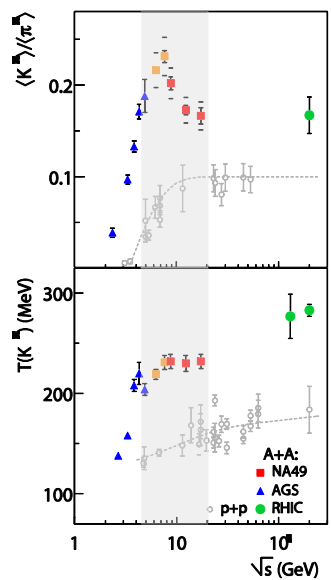

*Fig.1.* Collision energy dependence of the $K^+$-to $\pi^+$-mesons mean multiplicity ratio and the inverse slope parameter of the transverse mass spectra measured in central Pb+Pb and Au+Au collisions (solid symbols) compared to results from p+p reactions (open dots). The changes in the *SPS CERN* energy range (solid squares) suggest the onset of the deconfinement phase

Among all other results on ultrarelativistic heavy ion collisions concerning the states of the strongly interacting matter, the results which were obtained by M. Ga'zdzicki [1] were more attractive. The paper [1]} discussed that the results [2-3] on the energy dependence of hadron production in central *Pb+Pb* collisions at *20, 30, 40, 80* and *158 A GeV* coming from the energy scan program at the *CERN SPS* serve as evidence for the existence of a transition to the QGP [4].

Thus they are in agreement with the conjectures that at the top SPS CERN and *RHIC BNL* energies the matter created at the early stage of central *Pb+Pb* and *Au+Au* collisions is in the state of *QGP* [5-6]. The key results are summarized in Fig.1.

The most interesting effect can be seen in the energy dependence of the ratio $<K^+>/<\pi^+>$ of the mean multiplicities of $K^+$ and $\pi^+$, produced per event in the central *Pb+Pb* collisions, which is plotted in the top panel of the figure. Following a fast threshold rise, the ratio passes through a sharp maximum in the SPS CERN range and then seems to settle to a lower plateau value at higher energies. Kaons are the lightest strange hadrons and $<K^+>$ is equal to about half of the number of all anti-strange quarks produced in the collisions. Thus, the relative strangeness content of the produced matter passes through a sharp maximum at the *SPS CERN* in nucleus-nucleus collisions. This feature is not observed for proton-proton reactions. A second important result is the constant value of the apparent temperature of $K^+$ mesons in central *Pb+Pb* collisions at low *SPS CERN* energies as shown in the bottom panel of the figure. The plateau at the *SPS CERN* energies is preceded by a steep rise of the apparent temperature at the *AGS BNL* and followed by a further increase indicated by the *RHIC BNL* data.

Very different behavior is measured in proton-proton interactions. Presently, the sharp maximum and the following plateau in the energy of the $<K^+>/<\pi^+>$ ratio has only been reproduced by the statistical dependence model of the early stage[4] in which a first order phase transition is assumed. In this model the maximum reflects the decrease in the number ratio of strange to non-strange degrees of freedom and changes in their masses when deconfinement sets in. Moreover, the observed steepening of the increase in pion production is consistent with the expected excitation of the quark and gluon degrees of freedom. Finally, in the picture of the expanding fireball, the apparent temperature is related to the thermal motion of the particles and their collective expansion velocity. Collective expansion effects are expected to be important only in heavy ion collisions as they result from the pressure generated in the dense interacting matter. The stationary value of the apparent temperature of $K^+$ mesons may thus indicate an approximate constancy of the early stage temperature and pressure in the *SPS CERN* energy range due to the coexistence of hadronic and deconfined phases, as in the case of the first order phase transition [7-8].

Thus, the anomalies in the energy dependence of hadron production in central $Pb+Pb$ collisions at the low *SPS CERN* energies serve as evidence for the onset of deconfinement and the existence of *QGP* in nature. They are consistent with the hypotheses that the observed transition is of the first order. The anomalies are not observed in *p+p* interactions and they are not reproduced within hadronic models [9].

These results and their interpretation raise questions which can be answered only by new measurements. The energy region covered by the future measurements at the *SPS CERN* is indicated by the gray band.

**THE RESULTS OF THE EXPERIMENTS AT VARIOUS CENTRALITIES**

What about the results coming from the experiments at various centralities? Could these experiment indicate the effects same with the Ga'zdzicki's key results? Let us now consider some results from these experiments.

During the last several years, some results of the experiments at various centralities (see for example [10]) are discussed. These results demonstrate the point of regime change and saturation on the behavior of some characteristics of the events as a function of the centrality. We believe that such phenomena connected with fundamental properties of the strongly interacting matter could reflect the changes of its states (phases).

In [11] the variations of average transverse mass of identified hadrons with charge multiplicity have been studied for *AGS BNL*, *SPS CERN* and *RHIC BNL* energies (Fig.2). A plateau was observed in the average transverse mass for multiplicities corresponding to *SPS* energies. It was claimed that it can be attributed to the formation of a coexistence phase of quark gluon plasma and hadrons. So one can say that the central experiments confirm the existence of the plateau for the behaviors of *K*-mesons' temperature as a function of collisions centrality at the *SPS CERN* energies - the second key result of Ga'zdzidcki[1].

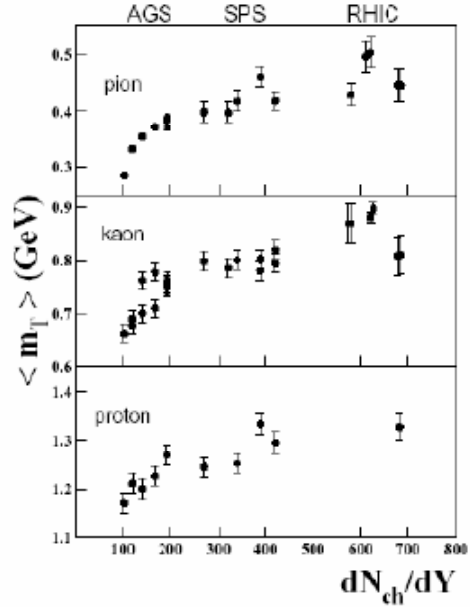

*Fig.2.* Variation of $<m_T>$ with produced charged particles per unit rapidity at mid rapidity for central collisions corresponding to different $\sqrt{s}$ spanning from *AGS BNL* to *RHIC BNL*.

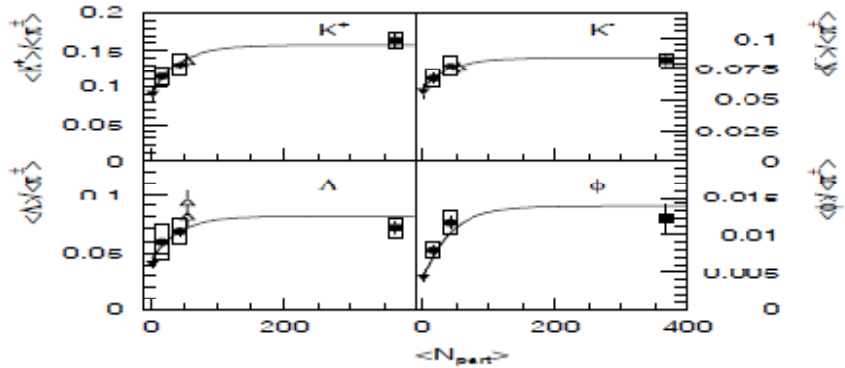

*Fig. 3.* The experimental ratios of the average values of multiplicity of $K^+$, $K^-$, $\varphi$-mesons and $\Lambda$-hyperons to the average values of multiplicity of $\pi^\pm$-mesons as a function of centrality ($<N_{part}>$).

Emission of $\pi^\pm$, $K^\pm$, $\varphi$ and $\Lambda$ was measured in near-central *C+C* and *Si+Si* collisions at *158 AGeV* beam energy[12]. Together with earlier data for *p+p*, *S+S* and *Pb+Pb*, the system-size dependence of relative strangeness production in nucleus-nucleus collisions are shown in Fig.3. Its fast rise and the saturation observed at about *60* participating nucleons can be understood as the onset of the formation of coherent systems of increasing size. So we could see that the results coming from the central experiments confirmed a fast threshold rise at *AGS BNL* energy range. But these results could not indicate any sharp maximum in the SPS CERN range.

So we could say that:

- the experiments at various centralities confirm the existing plateau for the behaviors of *K*-mesons' temperature as a function of collisions centrality at the SPS CERN energies - the second key result of Ga'zdzidcki[1];
- these experiments could indicate the increasing of the ratio $<K^+>/<\pi^+>$ at *AGS BNL* energies - the first key result of Ga'zdzidcki, but could not show the sharp maximum in the *SPS CERN* range.

We think that the last result could be connected with poorness of the experimental data for the ratio $<K^+>/<\pi^+>$ around of the values $N_{part} \cong 60$. This area could be investigated by increasing the intensity at the points using *NICA/MPD* [13] and *CBM* [14] experiments.

One remark is important that the definition of the centrality is not simple problem because it cannot be defined directly in the experiment. In different experiments the values of the centrality are fixed by different ways. The best way could be to select the events with a maximum number of nucleons - participants in the interaction. To do it, the following criteria are usually used: a number of identified protons, projectiles' and targets' fragments, slow particles, all particles, as the energy flow of the particles with emission angles $\theta \cong 0^0$ or with $\theta \cong 90^0$. Apparently, it is not simple to compare quantitatively the results on centrality-dependences obtained in literature while on the other hand the definition of centrality could significantly influence the final results. May be this is a reason, why we could not get a clear signal on new phases of strongly interacting matter. That is why it is very important to study the properties of the central collisions and to create some universal criteria to select those events to compare the result coming from different centrality experiments.

## NUCLEAR TRANSPARENCY EFFECT TO SEARCH FOR A SIGNAL ON NEW PHASES OF STRONGLY INTERACTING MATTER

New phases of strongly interacting matter (information on the *QCD* critical point) could be identified using the nuclear transparency effect - the behavior of $R = a_{AA}/a_{NN}$ ($a_{AA}$ and $a_{NN}$ are some values of the measured variables in the nuclear-nuclear and proton-proton experiments consequently, for example multiplicity of the charged particles), function at different energies as a function of the centrality. Because the transparency capability of different states of nuclear matter must be different. The investigations could give information on the onset state of the deconfinement as well as on *QCD* critical point.

Using data coming from codes and experimental data on the behavior of *R* as a function of the centrality it is possible to get information on the appearance of the anomalous nuclear transparency as a signal on changing of the states of strongly interacting matter. Nuclear transparency is one of the effects of nuclear-nuclear collisions from which one may obtain the information about the structure (at low energies), states (at middle energies), properties and phases of the nuclear matter (at relativistic and ultrarelativistic energies). The stated effect is a promising observable to map the transition between the different states/phases of the nuclear medium to the propagation of hadrons.

Transparency depends upon different factors of the events and the collisions:

(1) Impact parameter of the colliding particles.
(2) Geometry of projectile and target.
(3) Mass and size of the projectile and target.
(4) Diffuseness and density distribution of the projectile and target.
(5) Energies of the colliding particles.
(6) Atomic mass number of the target.
(7) Momentum transfer squared of the projectile.

Now one need to address the important question that how to fix the centrality and how one can study the centrality dependency of the *R* in the experiment?

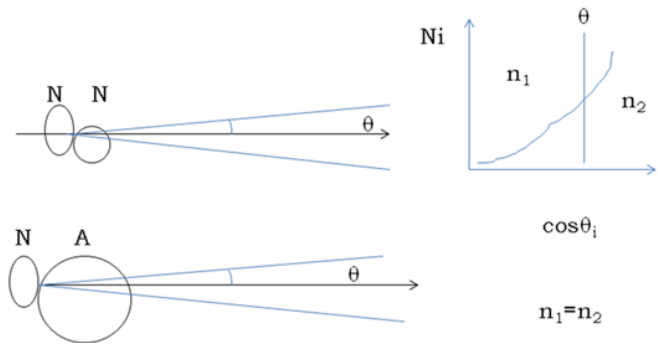

*Fig. 4.*

In papers [15] the nuclear transparency was used for the analysis of the data coming from the hadron- nuclear experiments. For this aim the authors used the inner cone term which was defined from the *NN* interaction (see Fig.4). The average values of multiplicity of fast particles - *s*-particles ($n_s$) were defined in the cone with half angle. Then the average values of *s*-particles with emitted angle less then $\theta$ were defined in nucleon-nuclear interactions (*NA*) and studied as a function of the number of the heavy particles - *h*-particles ($N_F$) emitted in the interactions. The values of the $N_F$ are used to fix the centrality. So there is some possibility to study the properties of the inner cone and outer cone particles produced in an event as a function of the centrality. In this paper they observed that the average multiplicity of inner cone particles did not depend on the centrality. It could be explained as a transparency of the nuclear matter for the inner particles emitting. The average values of the multiplicity for the outer cone particles increased linearly with $N_F$. We think that in the boards of different phases of strongly interacting matter the transparency don't change linearly with centrality. This result could be analyzed as a signal on phase transition in the strongly interacting matter. In [16], the effect of "transparency" of nuclear matter in interactions between $\pi^-$ - mesons and carbon nuclei was investigated at $P_\pi = 40\ GeV/c$. The following are their findings: for all chosen values of the limiting emission angle $\theta$ ($2.5^0$, $8^0$ and $10^0$) the average multiplicity of the $\pi^-$ - mesons of the inner cone does not depend on the number of emitted protons ($N_p$), and for $\theta = 2.5^0$ and $8^0$ it coincides with the results for the $\pi^-$ interactions; the fact that the average multiplicity of the $\pi^+$ - mesons of the inner cone is independent of the $N_p$ does not mean total "transparency" of the nucleus to these particles, since their average energy decreases

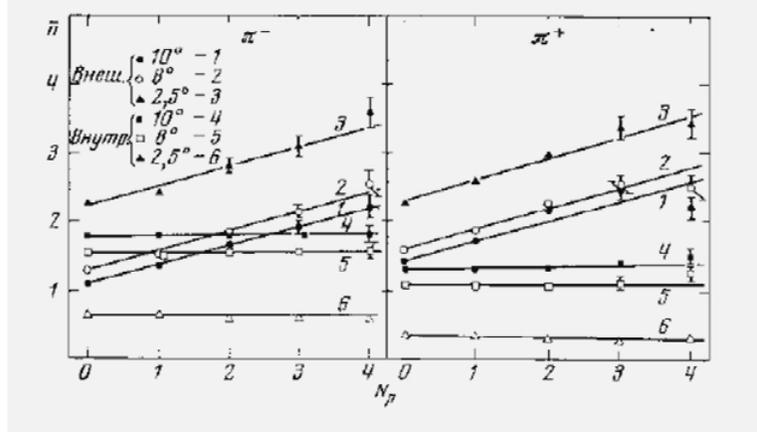

*Fig. 5.* The average values of multiplicity for $\pi^-$ (left panel) and $\pi^+$ (right panel) - mesons as a function of a number for identified protons in $\pi^{12}C$-reactions (lines were drift by hand)

**CONCLUSION**

The results coming from the central experiments confirm the existing saturation for the behaviors of $K$-mesons' temperature as a function of collisions centrality at the SPS energies.

These experiments could indicate the increasing of ratio $<K^+>/<\pi^+>$ at *AGS* energies, but could not show the sharp maximum in the *SPS* range. The result could be connected with insufficiency of the experimental points in the region of sharp maximum ($N_{part} \cong 60$).

*NICA/MPD* and *FAIR* experiments could get the necessary data to cover the region $N_{part} \cong 60$.

Study of the nuclear transparency effect as a function of the centrality could give an important information of the phases of strongly interacting matter.

We offer to use the inner cone definition to study the nuclear transparency effect as a function of the centrality.